# Alpha-Concave Hull, a Generalization of Convex Hull

Saeed Asaeedi[*,1], Farzad Didehvar[†,1], and Ali Mohades[‡,1]

[1]Department of Mathematics and Computer Science, Amirkabir University of Technology

**Abstract**

Bounding hull, such as convex hull, concave hull, alpha shapes etc. has vast applications in different areas especially in computational geometry. Alpha shape and concave hull are generalizations of convex hull. Unlike the convex hull, they construct non-convex enclosure on a set of points. In this paper, we introduce another generalization of convex hull, named alpha-concave hull, and compare this concept with convex hull and alpha shape. We show that the alpha-concave hull is also a generalization of an NP-complete problem named min-area TSP. We prove that computing the alpha-concave hull is NP-hard on a set of points.

## 1 Introduction

The computation of the region occupied by a set of points has been considered for many years. Computing the convex hull of a set of points is a way to represent the region occupied by the points. The Convex hull of a set of points is convex polygon with the minimum area that includes all these points. Many algorithms have been proposed to compute the convex hull [1–8]. In addition, convex hull is widely used in various fields such as shape matching [9], pattern recognition [10], image processing [11], fingerprint matching [12], geographical information systems [13, 14], path planning [15] etc.

The convex hull does not always specify the points region, accurately. There are many examples that the alpha shape represents the points region more accurately than convex hull. In [16], the alpha shape was originally introduced by Edelsbrunner. The alpha shape of a points set is a sub graph of Delaunay triangulation of points such that two points are connected if there is an empty ball of radius 1/alpha touching two points. When alpha=0, the ball of radius 1/alpha is replaced by half-plane, and hence the alpha shape of points will be equal to the convex hull of these points. So, the alpha shape is a generalization of convex hull such that it is applied in various fields such as pattern recognition [17, 18], bioinformatics [19], cosmology [20], sensor networks [21, 22] etc.

Another generalization of convex hull is the concave hull that has been introduced in [23] (Known as non-convex footprints) and then developed in [24]. The Concave hull of a set of points is an enclosure of the points by generating the non-convex polygons. Concave hull represents the tighter area occupied by the points than convex hull. Galton and Duckham suggested the Swing Arm algorithm based on gift wrapping algorithm [23], and Adriano and Yasmina suggested an algorithm based on the k-nearest neighbors approach [24] to compute the concave hull. In [25] an algorithm is presented to compute concave hull in n-dimension.

In this paper, we introduce a new generalization of convex hull, named Alpha-Concave Hull, to compute the region occupied by a set of points. The alpha-concave hull of a set of points, ACH, has following attributes: (i) ACH is a simple polygon, (ii) ACH includes all points, (iii) all internal angles of ACH are less than 180+alpha and (iv) the area of ACH is minimal. Computing alpha-concave hull for alpha=0 is equal to computing convex hull and for alpha=180 is equal to min-area TSP [26]. The formal definition of alpha-concave hull is exposed in section 2.

The document is organized as follows. In section 2, the concept of alpha-concave hull is introduced and some lemmas and theorems on this concept are proved. In the last section, we showed that alpha-concave hull is applied to the fields of shape approximations and cover designing. Better results are obtained in comparison with applying the convex hull and alpha shape.

## 2 Alpha-Concave Hull

We, first define the alpha-polygon to define the alpha-concave hull on a set of points. Alpha-polygon is a simple polygon in which all internal angles are less than 180+alpha. In the following, we will state the precise definition of alpha-polygon concept and alpha-concave hull.

**Definition 2.1** *The simple polygon A is called **α-polygon** if all interior angles of A are less than or equal to* $180 + \alpha$ *degrees.*

**Definition 2.2** *The **α-concave hull** of a set P of*

[*]asaeedi@aut.ac.ir
[†]didehvar@aut.ac.ir
[‡]mohades@aut.ac.ir

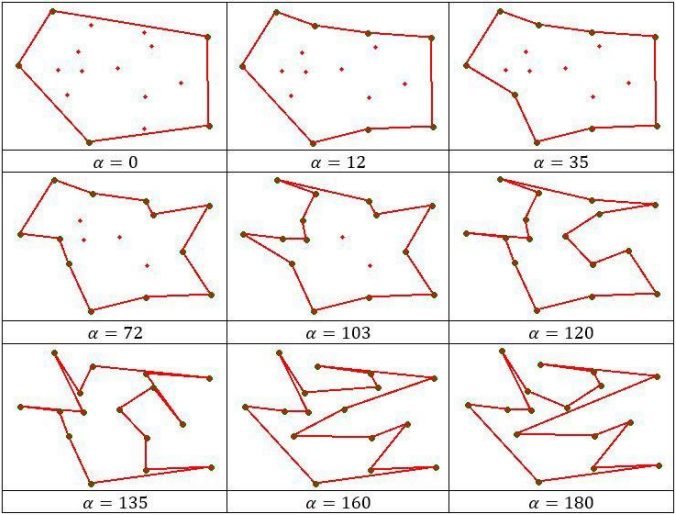

Figure 1: Alpha-concave hulls for a set of points.

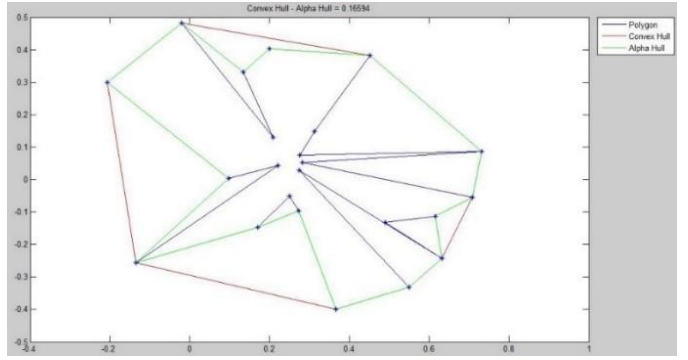

Figure 2: The blue line specifies the polygon. The red line specifies the convex hull. The green line specifies the alpha-concave hull.

points is the enclosing $\alpha$ − polygon with smallest area that contains P.

In definition 2.1, for $\alpha = 0$, the concept of $\alpha$-polygon is equal to the convex polygon. So, 0-concave hull of any set $P$ of points is equal to the convex hull of $P$. Consequently, the concept of alpha-concave hull is a generalization of convex hull. For $\alpha = 180$, the alpha-concave hull of $P$ is the minimal simple polygon consisting of $P$. The problem of computing the minimal simple polygon on a set of points is known as min-area TSP. So, the concept of alpha-concave hull is a generalization of min-area TSP. While computing min-area TSP is NP-complete, the computing of convex hull has an $O(nlogn)$ optimal algorithm.

Fig. 1 illustrates $\alpha$-concave hulls on a set of points for different values of alpha. The 0-concave hull is equal to the convex hull of points and the 12-concave hull is a semi-convex hull. For $\alpha > 100$, the $\alpha$-concave hulls construct sharp angles and the 180-concave hull is equal to the simple polygon with minimum area that contains all points.

The following theorems indicate the hardness of computing problem for an alpha-polygon with a given area on a set of points and the problem of computing alpha-concave hull.

**Theorem 1** *Computing $\alpha$ − polygon with a given area on the set of points is NP-complete.*

**Theorem 2** *Computing α-concave hull on the set of points is NP-hard.*

**Remark 1** *Fekete in [27] and [26] showed that finding the simple polygon with minimum area on a set of points, that is known as min-area TSP, is NP-complete.*

Since computing 180-concave hull is equal to the min-area TSP, we easily and more generally present the proof for this problem by theorem 2.

## 3 Application of alpha-concave hull

Since alpha-concave hull is a generalization of convex hull, we try to show that alpha-concave hull is generally more applicable than convex hull. Simultaneously, we compare the applications of alpha-concave hull to the same applications when we apply alpha shape.

Alpha-concave hull approximates the set of points or a polygon more accurately than convex hull. So, it enables us to apply alpha-concave hull instead of convex hull in shape approximation, cover designing, geometric modeling etc. Convex hull of points set $P$ is applied to approximate $P$ in path planning and path finding in the fields of robotics [15]. Applying alpha-concave hull gives us a better approximation of shapes rather than using convex hull. So, we expect that applying alpha-concave hull yield us superior maneuverability in robotics. The results in this section satisfy our expectation.

**Definition 3.1** *Let P be a polygon. We define **approximation error** of convex hull as the area between P and the convex hull of P and denote it by $E_{CH}(P)$.*

Clearly, alpha-concave hull is a subpolygon of convex hull for any polygon $P$. Hence, approximation error of alpha-concave hull is less than approximation error of convex hull. In other words, $E_{AlphaCH}(P) \leq E_{CH}(P)$. As illustrated in Fig. 2.

As stated before, the concept of alpha shape does not construct a convex polygon necessarily. Nevertheless, this concept is used for polygon approximation. To compare the approximation error of alpha-concave hull and alpha shape, we implement both approximations for 500 random scaled polygons. Table 1 reports the approximation errors of five samples of random poly-gons and the average error of approximations for 500 polygons.

Table 1: Approximation error of convex hull, alpha shape and alpha-concave hull

| # | Area | | | | Approximation Error | | |
|---|---|---|---|---|---|---|---|
| | Polygon | Convex Hull | Alpha Shape | α-concave hull | Convex Hull | Alpha Shape | α-concave hull |
| 1 | 0.146599413 | 0.284458498 | 0.584730356 | 0.284458498 | 0.137859085 | 0.438130943 | 0.137859085 |
| 2 | 0.156271926 | 0.487516049 | 0.487516049 | 0.32824526 | 0.331244123 | 0.331244123 | 0.171973334 |
| 3 | 0.285070702 | 0.435827316 | 0.365505776 | 0.365505776 | 0.150756615 | 0.080435074 | 0.080435074 |
| 4 | 0.13411634 | 0.314397968 | 0.288138401 | 0.293174548 | 0.180281628 | 0.154022061 | 0.159058208 |
| 5 | 0.337602248 | 0.600940344 | 0.535038758 | 0.546185385 | 0.263338096 | 0.197436509 | 0.208583137 |
| Ave | 0.242416363 | 0.455795892 | 0.402814933 | 0.400181389 | 0.213379529 | 0.160398569 | 0.157765026 |

Table 2: Performances of alpha shape and alpha-concave hull

| | α-concave hull < Alpha Shape | α-concave hull = Alpha Shape | α-concave hull > Alpha Shape |
|---|---|---|---|
| Count | 247 | 130 | 123 |

As shown in table 1, the average error of alpha-concave hull approximation is less than the average error of alpha shape approximation. Table 2 confirms that in 247 cases of polygon approximations, using alpha-concave hull provides more accurate results rather than using alpha shape.

In alpha shape approximation of polygon $P$, we found the best value of alpha which is a negative real number such that the approximated polygon $P'$ satisfied the following requirements: (i) $P'$ is a connected polygon, (ii) $P'$ contains $P$ and (iii) $P'$ has the minimum area.

Incorrect values of alpha lead to depict three possible shapes: A non-polygon shape, disconnected polygons as a shape, and a polygon that has not a minimum area. Similarly, in alpha-concave hull approximation of polygon $P$, we found the best value of alpha which is the greatest one to construct minimum polygon that contains $P$.

Convex hull, alpha-concave hull and alpha shape can be computed on a set of points or a mesh. Igarashi et al. use the convex hull on obtained mesh from 3D scanning of an object to design a cover [28]. Lucieer et al. use alpha shape to decompose a shape into visually meaningful parts [29]. Fig. 3 illustrates the fitness of alpha-concave hull against convex hull to design a cover.

In addition to the mentioned applications such as shape approximation, cover designing and path finding, the alpha-concave hull is applicable in pattern recognition fields including point pattern matching, fingerprint matching and digit recognition. Wen and Guo in [12] use the convex hull to eliminate spurious matching in fingerprint matching. Goshtasby and Stockman

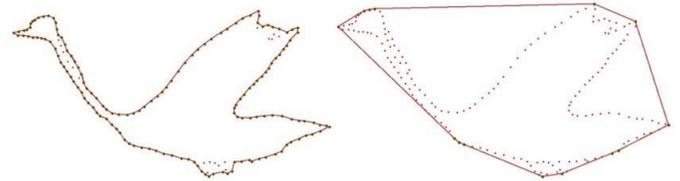

Figure 3: The result of alpha-concave hull algorithm on a meaningful set of points compared to the convex hull of the set.

use the convex hull in point pattern matching to reduce the search domain [30].

## 4 Conclusion

Summarily, we define a generalization of convex hull called α-concave hull such that the parameter α describes the smoothness level of computed hull on the points. When $α = 0$, alpha-concave hull is equal to convex hull. Loosely speaking, when $α = E$ is small enough; the alpha-concave hull is near to the convex hull. When $α = 180$, alpha-concave hull is the polygon of the minimum area crossing all vertices that is known as min-area TSP. We prove that computing alpha-concave hull is NP-hard. We implement this concept on a data set of random polygons. The results are more efficient and accurate comparing to applying convex hull and alpha shapes for the same data set.

## References

[1] Godfried T Toussaint. A historical note on convex hull finding algorithms. *Pattern Recognition Letters*, 3(1):21–28, 1985.

[2] Joseph o'Rourke. *Computational geometry in C.* Cambridge university press, 1998.

[3] Ronald L. Graham. An efficient algorith for determining the convex hull of a finite planar set. *Information processing letters*, 1(4):132–133, 1972.

[4] Donald R Chand and Sham S Kapur. An algorithm for convex polytopes. *Journal of the ACM (JACM)*, 17(1):78–86, 1970.

[5] Ray A Jarvis. On the identification of the convex hull of a finite set of points in the plane. *Information Processing Letters*, 2(1):18–21, 1973.

[6] William F Eddy. A new convex hull algorithm for planar sets. *ACM Transactions on Mathematical Software (TOMS)*, 3(4):398–403, 1977.

[7] Franco P. Preparata and Se June Hong. Convex hulls of finite sets of points in two and three dimensions. *Communications of the ACM*, 20(2):87–93, 1977.

[8] Michael Kallay. The complexity of incremental convex hull algorithms in $r_i \sup d_i$. *Information Processing Letters*, 19(4):197, 1984.

[9] Jonathan Corney, Heather Rea, Doug Clark, John Pritchard, Michael Breaks, and Roddy MacLeod. Coarse filters for shape matching. *Computer Graphics and Applications, IEEE*, 22(3):65–74, 2002.

[10] ZHANG Li Hua XU Wen Li. Convex hull based point pattern matching under perspective transformation. 2002.

[11] Zhengwei Yang and Fernand S Cohen. Image registration and object recognition using affine invariants and convex hulls. *Image Processing, IEEE Transactions on*, 8(7):934–946, 1999.

[12] Chengming Wen and Tiande Guo. An efficient algorithm for fingerprint matching based on convex hulls. In *Computational Intelligence and Natural Computing, 2009. CINC'09. International Conference on*, volume 1, pages 66–69. IEEE, 2009.

[13] Jiyuan Liu, Xinsheng Wang, and Dafang Zhuang. Application of convex hull in identifying the types of urban land expansion. *ACTA GEOGRAPHICA SINICA-CHINESE EDITION-*, 58(6):885–892, 2003.

[14] Ren-Can Peng, Jia-Yao Wang, Zhen Tian, Li-Xin Guo, and Zi-Peng Chen. A research for selecting baseline point of the territorial sea based on technique of the convex hull construction. *Cehui Xuebao/Acta Geodaetica et Cartographica Sinica*, 34(1):53–57, 2005.

[15] S Meeran and A Share. Optimum path planning using convex hull and local search heuristic algorithms. *Mechatronics*, 7(8):737–756, 1997.

[16] Herbert Edelsbrunner, David Kirkpatrick, and Raimund Seidel. On the shape of a set of points in the plane. *Information Theory, IEEE Transactions on*, 29(4):551–559, 1983.

[17] Matt Duckham, Lars Kulik, Mike Worboys, and Antony Galton. Efficient generation of simple polygons for characterizing the shape of a set of points in the plane. *Pattern Recognition*, 41(10):3224–3236, 2008.

[18] Marconi Soares Barbosa, Luciano da Fontoura Costa, and Esmerindo de Sousa Bernardes. Neuromorphometric characterization with shape functionals. *Physical Review E*, 67(6):061910, 2003.

[19] Weiqiang Zhou and Hong Yan. A discriminatory function for prediction of protein–dna interactions based on alpha shape modeling. *Bioinformatics*, 26(20):2541–2548, 2010.

[20] Rien van de Weygaert, Erwin Platen, Gert Vegter, Bob Eldering, and Nico Kruithof. Alpha shape topology of the cosmic web. In *Voronoi Diagrams in Science and Engineering (ISVD), 2010 International Symposium on*, pages 224–234. IEEE, 2010.

[21] Richard K Martin. Using alpha shapes to approximate signal strength based positioning performance. *Signal Processing Letters, IEEE*, 18(12):741–744, 2011.

[22] Marwan Fayed and Hussein T Mouftah. Localised alpha-shape computations for boundary recognition in sensor networks. *Ad Hoc Networks*, 7(6):1259–1269, 2009.

[23] Antony Galton and Matt Duckham. What is the region occupied by a set of points? In *Geographic Information Science*, pages 81–98. Springer, 2006.

[24] Adriano Moreira and Maribel Yasmina Santos. Concave hull: A k-nearest neighbours approach for the computation of the region occupied by a set of points. 2007.

[25] Jin-Seo Park and Se-Jong Oh. A new concave hull algorithm and concaveness measure for n-dimensional datasets. *Journal of information science and engineering*, 29(2):379–392, 2013.

[26] Sándor P Fekete and William R Pulleyblank. Area optimization of simple polygons. In *Proceedings of the ninth annual symposium on Computational geometry*, pages 173–182. ACM, 1993.

[27] Sándor P Fekete. On simple polygonalizations with optimal area. *Discrete & Computational Geometry*, 23(1):73–110, 2000.

[28] Yuki Igarashi and Hiromasa Suzuki. Cover geometry design using multiple convex hulls. *Computer-Aided Design*, 43(9):1154–1162, 2011.

[29] Yanyan Lu, Jyh-Ming Lien, Mukulika Ghosh, and Nancy M Amato. $α$-decomposition of polygons. *Computers & Graphics*, 36(5):466–476, 2012.

[30] Ardeshir Goshtasby and George C Stockman. Point pattern matching using convex hull edges. *Systems, Man and Cybernetics, IEEE Transactions on*, (5):631–637, 1985.